\title{A Computational Approach for Checking Compliance with European View and Sunlight Exposure Criteria}
\author{Eleonora Brembilla, Shervin Azadi, Pirouz Nourian}

\documentclass[conference]{IEEEtran}
\def\BibTeX{{\rm B\kern-.05em{\sc i\kern-.025em b}\kern-.08em
    T\kern-.1667em\lower.7ex\hbox{E}\kern-.125emX}}

\usepackage{microtype}

\usepackage{times}

\usepackage{fancyhdr,graphicx,amsmath,amssymb, amsfonts,mathtools}
\usepackage{euscript} %euler script
\usepackage[table]{xcolor}% http://ctan.org/pkg/xcolor
\usepackage[
	linesnumbered,
	boxed,
	ruled,
	vlined
]{algorithm2e}
\usepackage{hyperref}
\hypersetup{
    colorlinks=true,
    linkcolor=blue,
    filecolor=magenta,      
    urlcolor=cyan,
}
\SetCommentSty{mycommfont}
\SetNlSty{mynumfont}{}{}
\SetAlCapFnt{\footnotesize}
\SetKwFor{While}{while}{do}{end}
\usepackage{setspace}

\usepackage{enumitem}% http://ctan.org/pkg/enumitem
\let\OLDitemize\itemize
\renewcommand\itemize{\OLDitemize[noitemsep,topsep=0pt]}
\let\OLDenumerate\enumerate
\renewcommand\enumerate{\OLDenumerate[noitemsep,topsep=0pt]}

\usepackage{multirow,tabularx, booktabs} % advanced table package
\usepackage[justification=centering, 
font=small, 
labelfont=bf, 
textfont=it]{caption} % package for adding and styling the captions

\usepackage{float} % to affect the placement of elements
\usepackage{placeins} % to place barriers for float objects
\usepackage{subcaption} % for captioning subfigures

\usepackage{hyperref} % package for refrencing the tables, figures and bibliography
\usepackage[
    backend=biber,
    style=ieee
]{biblatex} % bibliography package and styling

\newcommand{\comment}[1]{} % multiline comment function

\usepackage{algorithmic}
\usepackage{textcomp}
\hypersetup{
	pdftitle={A Computational Approach for Checking Compliance with European View and Sunlight Exposure Criteria},
	pdfauthor={Eleonora Brembilla, Shervin Azadi, Pirouz Nourian},
}

\begin{document}
\title{A Computational Approach for Checking Compliance with European View and Sunlight Exposure Criteria
}

% \author{
% 	\IEEEauthorblockN{Eleonora Brembilla}
% 	\IEEEauthorblockA{\textit{dept. name of organization (of Aff.)} \\
% 	\textit{Delft University of Technology}\\
% 	Delft, the Netherlands \\
% 	email address or ORCID}
% 	\and
% 	\IEEEauthorblockN{Shervin Azadi}
% 	\IEEEauthorblockA{\textit{dept. name of organization (of Aff.)} \\
% 	\textit{Delft University of Technology}\\
% 	Delft, the Netherlands \\
% 	email address or ORCID}
% 	\and
% 	\IEEEauthorblockN{Pirouz Nourian}
% 	\IEEEauthorblockA{\textit{dept. name of organization (of Aff.)} \\
% 	\textit{Delft University of Technology}\\
% 	City, Country \\
% 	email address or ORCID}
% }

% \author{
% 	\IEEEauthorblockN{Eleonora Brembilla}
% 	\IEEEauthorblockA{\textit{Architectural Engineering \& Technology} \\
% 	\textit{Delft University of Technology}\\
% 	Delft, the Netherlands \\
% 	0000-0003-1800-8660}
% 	\and
% 	\IEEEauthorblockN{Shervin Azadi}
% 	\IEEEauthorblockA{\textit{Architectural Engineering \& Technology} \\
% 	\textit{Delft University of Technology}\\
% 	Delft, the Netherlands \\
% 	0000-0002-8610-2774}
% 	\and
% 	\IEEEauthorblockN{Pirouz Nourian}
% 	\IEEEauthorblockA{\textit{Architectural Engineering \& Technology} \\
% 	\textit{Delft University of Technology}\\
% 	Delft, the Netherlands \\
% 	0000-0002-3817-7931}
% }

\begin{author}
	\author{\author{
		\IEEEauthorblockN{Eleonora Brembilla, Shervin Azadi, and Pirouz Nourian}
		\IEEEauthorblockA{Department of Architectural Engineering \& Technology,}
		\textit{Delft University of Technology}\\
		Email: \{e.brembilla, , s.azadi-1, p.nourian\}@tudelft.nl
	}}
\end{author}

\maketitle
\thispagestyle{plain}
\pagestyle{plain}
\begin{abstract}
    The paper presents open-source computational workflows for assessing the ``Exposure to sunlight'' and ``View out'' criteria as defined in the European standard EN 17037 ``Daylight in Buildings'', issued by the European Committee for Standardization. In addition to these factors, the standard document also addresses daylight provision and protection from glare, both of which fall out of the scope of this paper. The purpose of the standard is stated as `encouraging building designers to assess and ensure successfully daylit spaces'. The standard document proposes verification methods for performing such assessments, albeit without  %aimed at determining the ranges of ``exposure to sunlight" and the quality-levels of ``view out"; % marked with the labels ``minimum, medium, and high". The standard document proposes normative methods for ensuring minimal quality levels pertaining to the geometry of buildings; 
recommending a simulation procedure for computing the aforementioned criteria. The workflows proposed in this paper are arguably the first attempt to standardize these assessment methods using de-facto open-source standard technologies currently used in practice. The approach of this work is twofold: establish that the compliance check can be systematically performed on a 3D model by a novel simulation tool developed by the authors; and highlighting the additional assumptions that need to be implemented to build a robust and unambiguous tool within existing open-source frameworks. \footnote{Code available at \url{https://github.com/shervinazadi/EN_17037_Compliance}}
%The purpose of this endeavour is twofold: on the one hand, the paper is to reflect on the issues pertaining to the computational standardization of such standards, and on the other hand, it is to pave the way for further open-science research in the field of building simulation utilizing state of the art open-source scientific computing technologies. 

\end{abstract}

\section*{Key Innovations} % 1-5 bullet points

\begin{itemize}
	\item Formulating procedures for computational assessment of EN 17037 criteria for sunlight exposure and view out
	\item Standardizing inputs and outputs for the proposed computational assessment procedures
	\item Devising Python workflows utilizing the Radiance simulation engine and Honeybee from Ladybug Tools for running sunlight exposure and view analyses
\end{itemize}

\section*{Practical Implications} % 20-50 words
The tool presented in this paper offers an open-source solution to check compliance with EN 17037 criteria for ``View out'' and ``Exposure to sunlight''. The tool will be made available to designers, consultants and researchers. Furthermore, the paper presents suggestions to policy-makers on how the compliance procedures could be made more robust and better integrated into computational design workflows.
%The main practical purpose of the paper is to make such quality assessments as ``sunlight exposure'' and ``view out'' openly accessible and digitally reproducible. The proposed workflows run light simulations using the so-called `recipes' framework of Honeybee \cite{Roudsari2013a,Subramaniam2018} for utilizing Radiance engine in Jupyter interactive Python notebooks. While the exemplary standard document addressed in this paper sets some value levels to be attained, it does not propose an explicit way to test compliance with the defined levels. As such, the proposed workflows showcase the possibility of automatically checking compliance with such standards on an open platform using free and open-source tools. 

\section{Introduction}

In 2018 a new European standard on ``Daylight in Buildings'', EN 17037 \cite{CEN2018} was introduced and ratified in all European Union countries, largely superseding existing national daylighting standards. The document includes recommendations for: 1) Indoor daylight provision; (2) View out; (3) Exposure to sunlight; and (4) Protection from glare. Previous works focused mainly on the implementation of the ``Indoor daylight provision'' and ``Protection from glare'' criteria \cite{Jones2019,Sprah2019,Paule2019,Sepulveda2020}. Instead, the proposed paper focuses on the ``View out'' and ``Exposure to sunlight'' sections; namely, it introduces a computational approach to facilitate compliance assessment at early design stages.

The analyses required by these two sections of the standard are exclusively dependent on the geometrical characteristics of the building under assessment and its surroundings. The verification procedures suggested in the standard are largely based on geometrical measurements on 2D plan and section views. Separate options are given for the verification via photographs -- only for existing buildings -- or via rendered images. For the assessment of view quality, the theoretical basis and a description of the advanced verification method can be found in \cite{Hellinga2013} \cite{Hellinga2014}, respectively. As for sunlight exposure, the verification procedure is delineated by \cite{Darula2015} and is based on evidence of the benefits of direct sunlight in buildings found by \cite{Neeman1976}. 

% After the European standard has been published, the recommended performance indicators and targets went through further scrutiny. Recent research on the ``View out'' criterion explored the relationship between performance indicators and assessments on building occupants' preferences in existing buildings. \cite{Kuhlenengel2019} investigated relations between the view criteria and students' achievements in 220 US classrooms, but found a significant positive correlation only between the number of layers and reading achievements. \cite{Waczynska2020} assessed relations between the occupants' ability to quantify the view criteria with a subjective evaluation and the results obtained from the computation of the same criteria with a 3D model. They found that participants could reliably judge the number of `view layers' visible from a window, but not the `horizontal sight angle' value. It is worth mentioning that in these studies, given that the analysis was performed on existing buildings, all or parts of the variables required by the assessment procedure were manually measured in situ.

We argue that being able to assess all four criteria with a single computational tool would facilitate the uptake and application of the new standard from early design decision-making. Additionally, being able to assess the ``View out'' and ``Exposure to sunlight'' with procedures based on direct ray-tracing is deemed to be more efficient than generating rendered images, and better suited for computational design processes. The approach of this work is twofold: establish that the compliance check can be systematically performed on a 3D model by a novel simulation tool developed by the authors; and highlighting the additional assumptions that need to be implemented to build a robust and unambiguous tool within existing open-source frameworks.
%Three scenarios characterised by different levels of density and complexity are analysed. 
%The work aims at testing the feasibility computational assessment of the proposed criteria of the new standard compatible with the contemporary design practice -- characterized by the use of 3D models and parametric/procedural workflows –- \red{and that the recommended targets are achievable in realistic urban and rural scenarios.}

%\subsection{Problem Statement}

Given that the verification procedures proposed in the standard document are developed primarily for manual geometrical assessment, several challenges and problems surfaced in developing the equivalent simulation procedures. The present paper describes how we reinterpreted this standard for a computational assessment based on ray-tracing. The main problems addressed in this paper are the standardization of definitions, data models and procedures for performing the proposed assessments. Ideally, this should be carried out in such an unambiguous way that an algorithm can perform the assessment and put out a conclusion without the need for a human interpreter during intermediate steps.

\section{Methods}
In this section we present in detail the procedures devised to determine target levels recommended by sections 5.2 and 5.3 of EN 17037:2018, respectively addressing the assessments of ``View out'' and ``Exposure to sunlight''. \comment{Firstly the relevant parts of the EN 17037 standard were analysed in depth to understand which are the fundamental variables to be computed} The chosen performance indicators and recommended target values are listed in Table \ref{tab:indicators} and explained below.

\begin{table}[htbp]
	\captionsetup{font=small}
	\caption{Performance level indicators and recommended target values for \textit{view} and \textit{exposure to sunlight}.}
	\label{tab:indicators}
	\centering
		\begin{tabularx}{\columnwidth}{
			        >{\footnotesize\centering\arraybackslash\hsize=0.6\hsize}X
			        >{\footnotesize\centering\arraybackslash\hsize=1.1\hsize}X
			        >{\footnotesize\centering\arraybackslash\hsize=1.1\hsize}X
			        >{\footnotesize\centering\arraybackslash\hsize=1.1\hsize}X
			        >{\footnotesize\centering\arraybackslash\hsize=1.1\hsize}X
			    }
		\hline
		& \multicolumn{3\footnotesize}{c}{View out} & Exposure to sunlight \\
		\hline
		& Horizontal sight angle & Distance to obstructions & Number of layers & Sunlight hours \\
		\hline
		Min & $\geq$14 & $\geq$6 & 1 & 1,5 \\
		Med & $\geq$28 & $\geq$20 & 2 & 3 \\
		High & $\geq$54 & $\geq$50 & 3 & 4 \\
		\hline
	\end{tabularx}
\end{table}

The combination of the first three indicators should in principle provide a comprehensive assessment of the quality of the view out. The overall performance level (classified as minimum, medium or high) is equivalent to that of the indicator that scores the lowest among the three. The horizontal sight angle describes the width of a window as seen by an observer's point of view. The distance to obstructions is the distance between the interior surface of a window and the ``major'' outdoor obstructions to a view of the sky. The number of layers gives an indication of how varied the outside view is by considering the presence of at least \comment{was written as most} one of the three types of elements: a view of the ground, a view of the landscape (natural, architectural or of the horizon), and a view of the sky. The fourth indicator from Table \ref{tab:indicators} refers to the exposure to sunlight recommendations and indicates the number of hours during which a point on the inner\comment{how does the inner/outer surface make a difference here?} the surface of a window receives direct sunlight.

Besides the imprecision inherent in the definition of some key factors in determining the quantity of these performance indicators, a number of variables are left for the user/designer to decide. Among these, two were identified as critical for understanding the variability of results: the view-point location for the view out analysis, and the date for the sunlight exposure analysis. The variability of performance indicators due to these two variables was investigated using the newly developed tools and is presented in the Result section. The following sections describe the computational approach and the test scene used for the analysis.

\subsection{Computational tools}

To maximize accessibility and ease of use, we adopted existing technical frameworks and built on top of them. The developed toolset is primarily available as a Python package. The simulation core utilizes the Radiance ray-tracer \cite{WardLarson1998b} through the utility functions and methods of the Ladybug Tools software \cite{Roudsari2013a,Subramaniam2018}. 

\subsubsection{View Out}

The ``Context View'' recipe assesses the view of a vantage point inside the room towards outside through a window, by computing the horizontal view angle, the distance to outdoor obstructions, and the view layers that are visible from that point. The standard does not constraint choosing the horizontal positioning of the vantage point (as long as it is in the so-called utilized area), but it limits the point height to 1.2\,m for a sitting eye level and 1.7\,m for a standing eye level. As in the case of sunlight exposure, the standard proposes two methods: one based on fish-eye photographs and one geometrical method. In the fish-eye method, it is suggested to superimpose a diagram on the photograph, in order to measure the horizontal angle and counting the different types of layer (sky, ground, landscape) present in the photograph. In the geometrical method, a straight line should be drawn from the vantage point towards the upper sill and lower sill in the section view and extended towards the outside scene until it hits an obstacle. This line represents the bounds of what is visible from the vantage point. Based on this drawing, the number of visible layers can be counted. The horizontal sight angle can be determined via a set of graphs relating the width of the window to the depth of the room.

To translate this procedure into a simulation workflow, we discretize the horizontal plane of the room at a height of 1.2\,m and 1.7\,m into a grid of vantage points. Then, from each vantage point, rays are shot towards a homogenous discretization (subdivided icosahedron) of a sphere, mapping the distance and type of objects in all directions. 

\begin{figure}[!htbp]
	\centering
	\includegraphics[width=\linewidth]{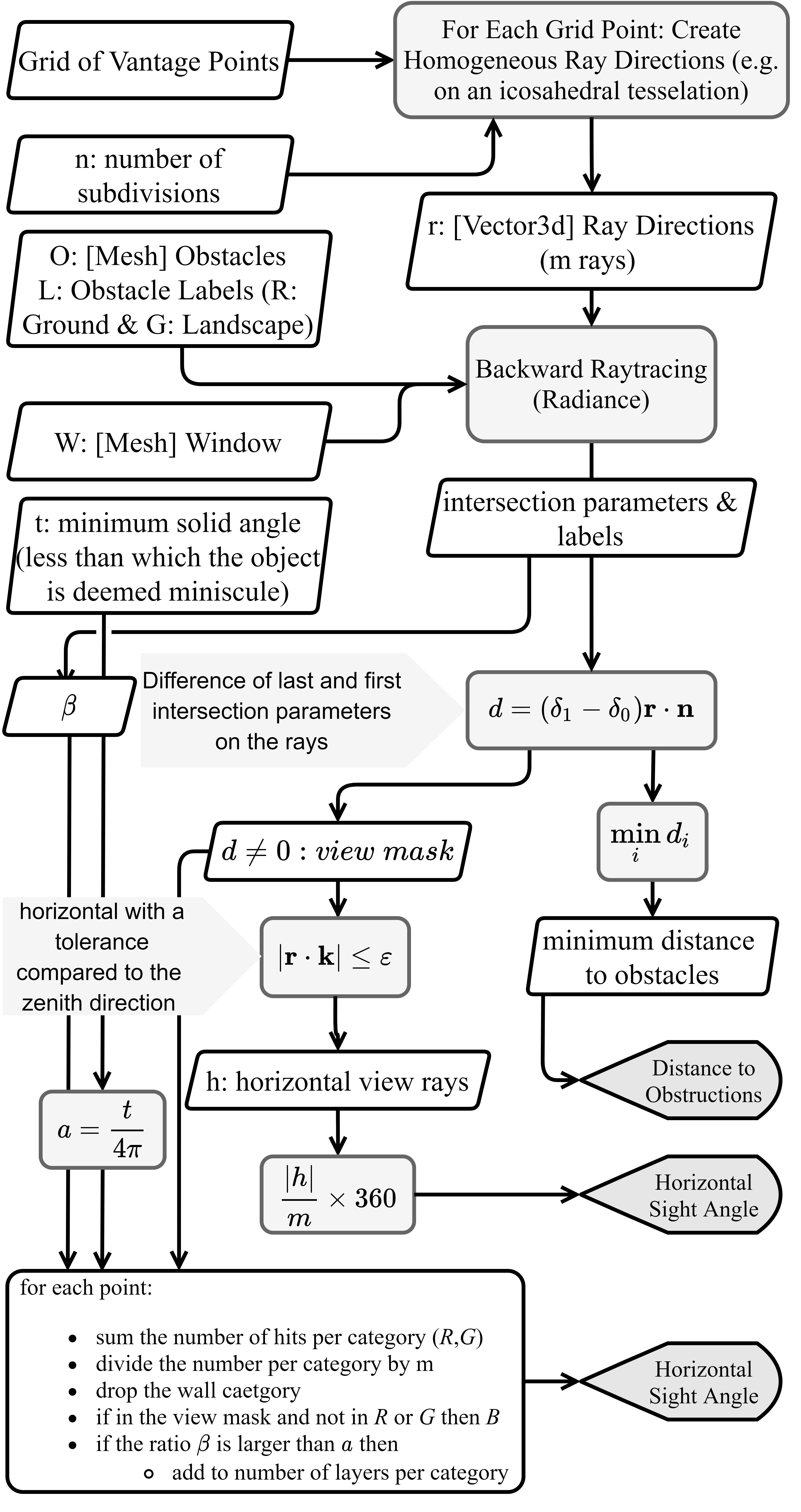}
	\caption{Context View simulation diagram.}
	\label{fig:viewdiagram}
	% \vspace{-4mm}
\end{figure}

%TODO correct dot product symbol

\textbf{Distance to obstructions}: For all shot rays, we retrieve the distance to the first and last intersection points. The difference between these two values enables us to identify the rays that have passed the window. Given that the standard specifies that the distance to outside obstructions should be measured perpendicular to the window surface, the distance can be computed as: 
\begin{equation}
	d = (\delta_1 - \delta_0) \mathbf{r} \cdot \mathbf{n} 
\end{equation}
where $\mathbf{n}$ is the window surface normal, $\mathbf{r}$ is the ray direction, $\delta_0$ is the distance of the vantage point to the first intersection, and $\delta_1$ is the distance of the vantage point to the last intersection.

As further elaborated in the discussion, the standard document does not specify the nature of the scene as a geo-spatial data model. We have pragmatically considered the maximum extent of our scene model (circa 500 metres) as the farthest intersection distance for the test case discussed here. However, it must be noted that without specifying the extents of the scene there might be a range of unforeseen issues arising out of such an ambiguity in the standard document, especially for a computational assessment. Namely, an arbitrary choice of such a distance will inevitably affect the reproducibility of the results and thus the reliability of the standard. Furthermore, in regions where the shape of the terrain is non-trivially blocking the sunlight hours (e.g. because of a mountain range behind a building, even if far away) then the results obtained as such cannot be reliable. At such a scale, i.e. major terrain obstacles, even the curvature of the earth will play a role. Past a certain distance, e.g. circa 8 km away from a 4000 meters high mountain peak, it will not be visible on the horizon due to the curvature of the earth. However, in closer ranges high mountainous ranges can potentially block the sunlight hours.
%TODO add longer explanation as per reviewer's comment

\textbf{Horizontal sight angle}: To compute the horizontal sight angle, we assess the ratio of the horizontal rays that have passed the window ($\rho$) to all the horizontal rays ($\varrho$), i.e.\ the portion of a horizontal circle covered by the window from a specific vantage point. Given that rays are shot toward a homogenous discretization of the sphere, the horizontal sight angle can be computed as the following:
\begin{equation}
	\beta =  360 \frac{\rho}{\varrho}
\end{equation}

\textbf{View layers}: On each grid point, we count the number of final ray intersections for each object type. Consequently, we can compute how many view layers were in the sight of each vantage point and what portion of the view sphere is occupied by them. If these portions are beyond a certain threshold, we can establish that the object is visible and perceivable by building occupants. The need for such a threshold and its specification will be further examined in the discussion section.

\subsubsection{Exposure to Sunlight}

The ``Exposure to Sunlight'' recipe is a measure of the minimum number of hours that sun is visible from a vantage point $P$ defined in the horizontal centre of the window with the minimum height of 1.2\,m above the floor or 0.3\,m above the sill. The standard specifies that the selected date for measurement should be between February 1st and March 21st. To compute this measure, two methods are proposed in the document: one based on a fish-eye photograph taken from the vantage point, and one based on geometrical calculations. In the fisheye method, a sun path diagram is superimposed on the photograph, so that the number of hours that directly see the sun can be counted. In the geometrical method, $\alpha$ is defined as the horizontal acceptance angle of the window spanning from the right side to the left side of the window from $P$, excluding major external obstacles. Next, the sun path for different geographical locations can be identified via a lookup table. Finally, with the help of the angle between window normal and North direction, the overlap between $\alpha$ and the sun path span is determined.

\begin{figure}[!htbp]
	\centering
	\includegraphics[width=\linewidth]{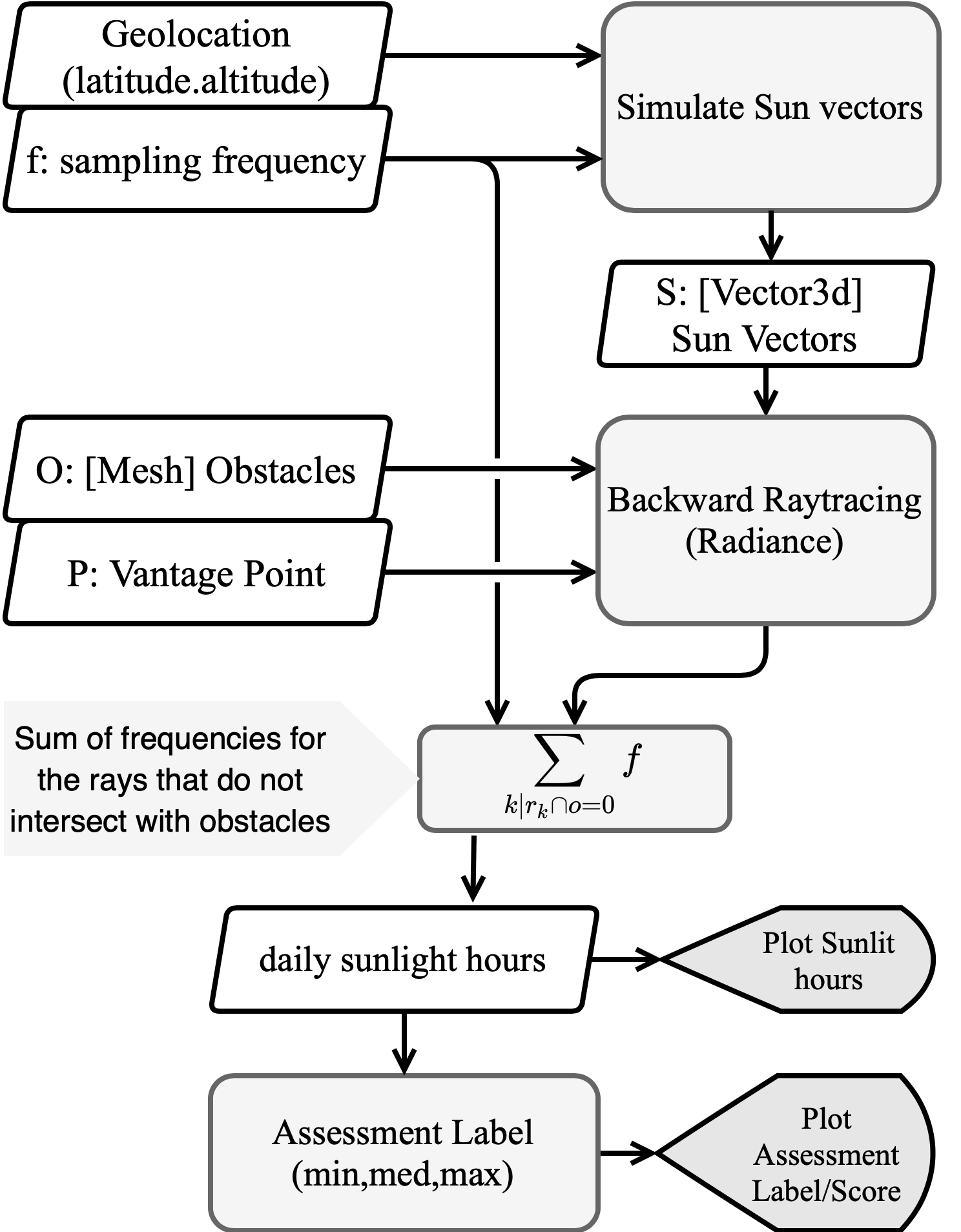}
	\caption{Exposure to Sunlight simulation diagram.}
	\label{fig:solardiagram}
	% \vspace{-4mm}
\end{figure}

It is clear that such assessment of the overlap of the sun path with the opening of the window is extremely sensitive to the precision of drawings and measurements. Moreover, the more complex the outdoor scene becomes, the less feasible is to compute the effect of obstructions with this procedure. Consequently, to translate this procedure into a simulation workflow (illustrated in Figure \ref{fig:solardiagram}), we computed the position of the sun throughout the year in half an hour intervals based on the geolocation of the building. Then, by considering the geometry of the context as obstacles we checked for a clear sight-line from the vantage point of the window ($P$) in the direction of the sun. Aggregating this information for each day indicates the number of sunlight exposure hours.

\subsection{Test Scene}
\begin{figure}[!htbp]
	\centering
	\includegraphics[width=\linewidth]{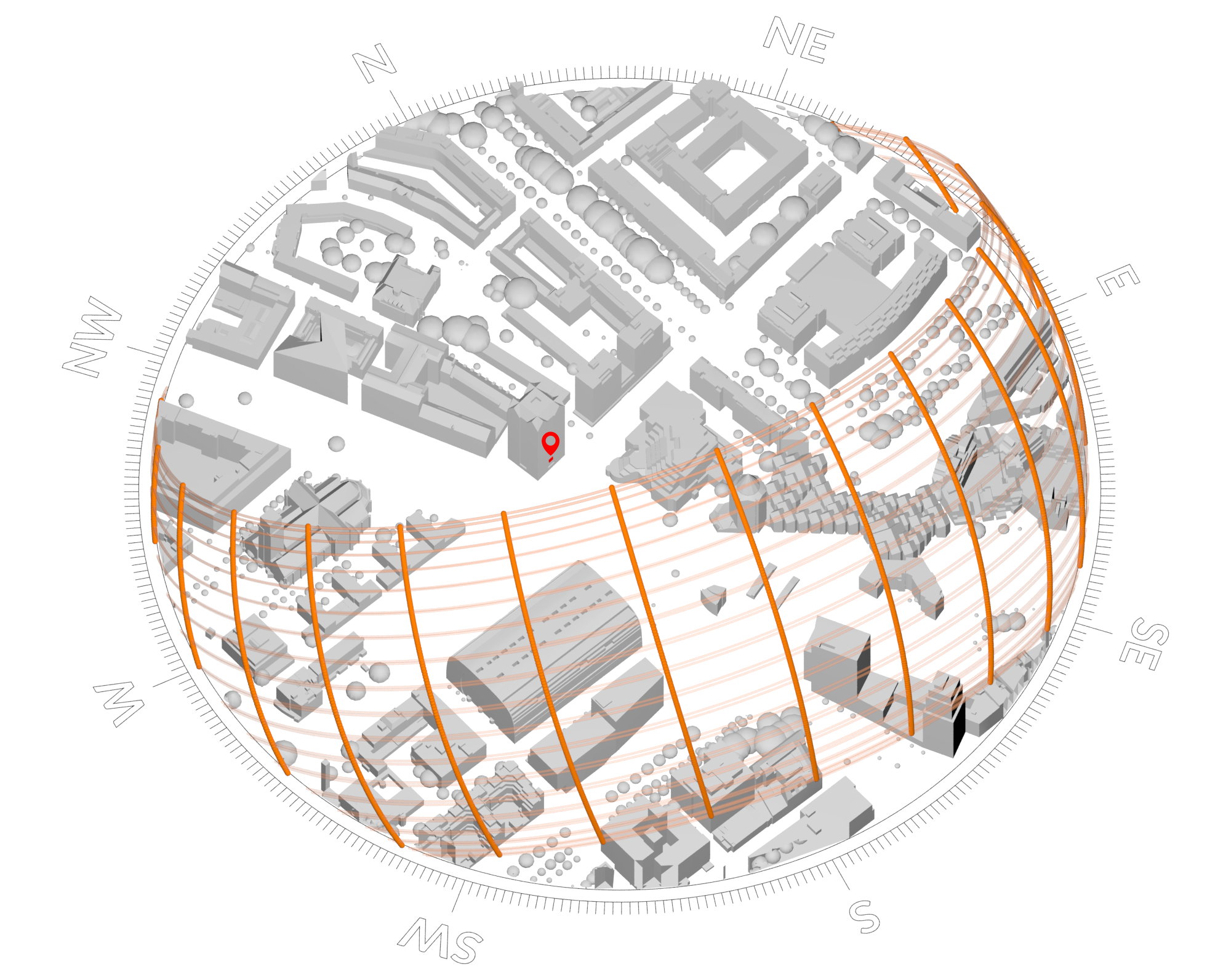}
	\caption{Bird view of the scene indicating the relative positioning of the test-case, in red; urban surrounding in light grey; hourly sun positions, as dark orange spheres; and bi-weekly sun paths as light orange lines.}
	\label{fig:birdview}
	% \vspace{-4mm}
\end{figure}

\begin{figure}[!htbp]
	\centering
	\includegraphics[width=\linewidth]{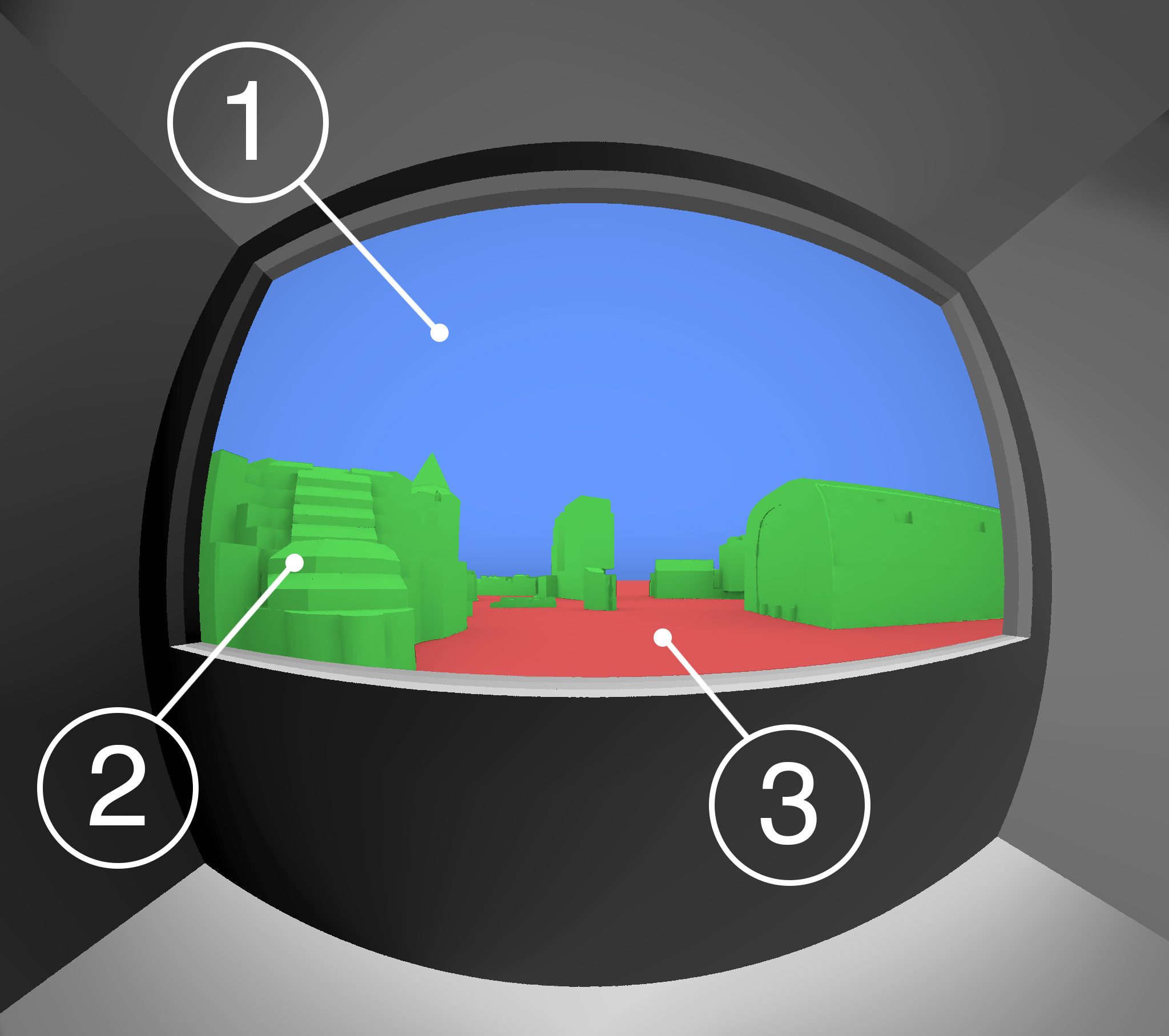}
	\caption{Outdoor view as seen from a point within the room, at a height of 1.2\,m above the floor. From this specific viewpoint, it is possible to see all the three `view layers': (1) Sky, in blue; (2) Landscape, in green (2); and (3) Ground, in red.}
	\label{fig:eyeview}
%	\vspace{-4mm}
\end{figure}

Two different 3D models were combined to recreate a fictitious -- yet realistic -- test scene for the evaluation. For the urban environment, an existing model of Rotterdam city centre was used \cite{rotterdam_rotterdam_nodate}. Such a model includes the Blaaktoren tower, where the Massachusetts Institute of Technology (MIT) reference office model was inserted, namely on the third floor. The office model represents a deep-plan, side-lit room used for benchmarking daylighting and glare analyses \cite{Reinhart2013}. Figure \ref{fig:birdview} shows the test scene and indicates the position of the office room, as well as the sun path diagram for the coordinates of Rotterdam city centre. The office window is oriented towards the South-East and overlooks the view represented in Figure \ref{fig:eyeview}. The `view layers' visible in this example are colour-coded in a similar way to \cite{Mardaljevic2020}.

%TODO check the computation for difference between solar and civil time and correct the solar path diagram in the figure

\section*{Results}

\subsection{View out}

The first set of results concerns the assessment of view quality. The performance criteria related to ``View Out'' (number of view layers and horizontal sight angle) were computed and visualized, for two grids of points: one at the height of 1.2\,m above the floor and the other at 1.7\,m. The first grid represents the view of a sitting person, whereas the second grid represents the view of a standing person. The third criterion to assess view quality -- distance to obstructions -- was calculated for multiple points on the window surface, as the standard does not define a specific point for this assessment. The minimum distance was found to be 16\,m and the maximum was 351\,m. If the minimum distance was to be considered, then the view performance of this room would result in a `Minimum' score. Oppositely, if the point chosen for the assessment would result in a distance of 351\,m, then the view would score a `High' performance level. The placement of the room in the urban model and the resulting variability in obstruction distances were purely coincidental. However, it is reasonable to expect that most rooms with a high number of view layers will be characterised by a large variety of distances to major obstructions. Hence, such an assessment would benefit from the definition of a single statistical indicator that can give a meaningful indication of the overall performance, for example, the median value of all computed distances.

\begin{figure}[!htbp]
	\centering
	\includegraphics[width=\linewidth]{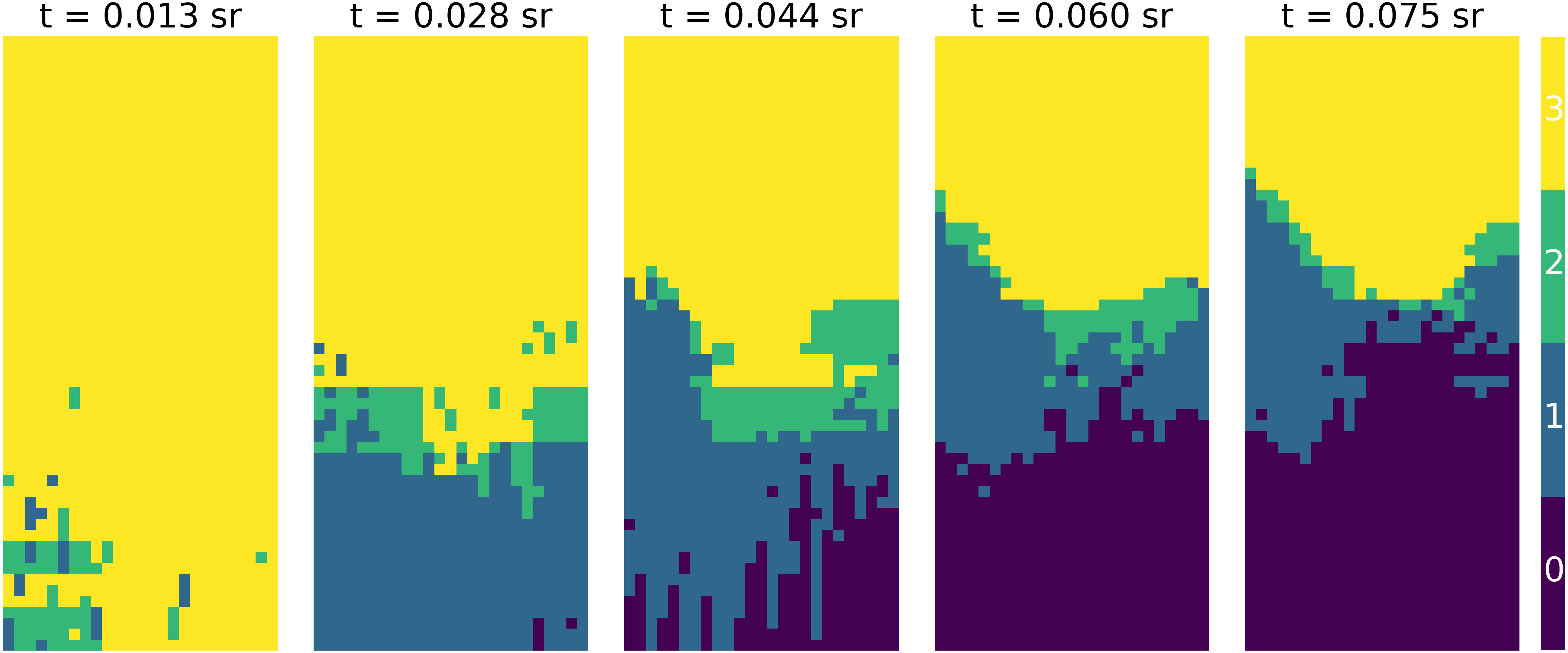}
	\caption{Number of view layers visible across a grid of analysis points 1.7\,m high. ($t$) is the visibility threshold for counting a layer as visible and is measured in steradians. The window is on the top side.}
	\label{fig:viewresult_layers}
	% \vspace{-4mm}
\end{figure}

Figure \ref{fig:viewresult_layers} shows the analysis on the number of view layers for a grid at 1.7\,m. From the first plot on the left, it looks as if from almost the entire room is possible to see all three view layers. For such a deep plan room, it is surprising that the ground can be seen even when standing at the furthest point from the window. What happens is that the count of view layer is equally sensitive to very large or very small portions of each view element. That is, even if only one ray hits, e.g., the ground, then such layer is included in the overall count. An additional investigation was therefore conducted on the effect of adding a minimum threshold value for a layer to be counted as effectively part of the view. The other plots in Figure \ref{fig:viewresult_layers} illustrate how the evaluation changes depending on the set minimum threshold. It is also possible to notice that the back portion of the room does not see any view layer. Although a view out of the window is always present, from those points none of the layers covers a solid angle exceeding the defined threshold. Thus, the introduction of such a threshold could be used to indicate when the view out is too distant from the selected vantage point, effectively removing the need to assess the horizontal sight angle.

\begin{figure}[!htbp]
	\centering
	\includegraphics[width=\linewidth]{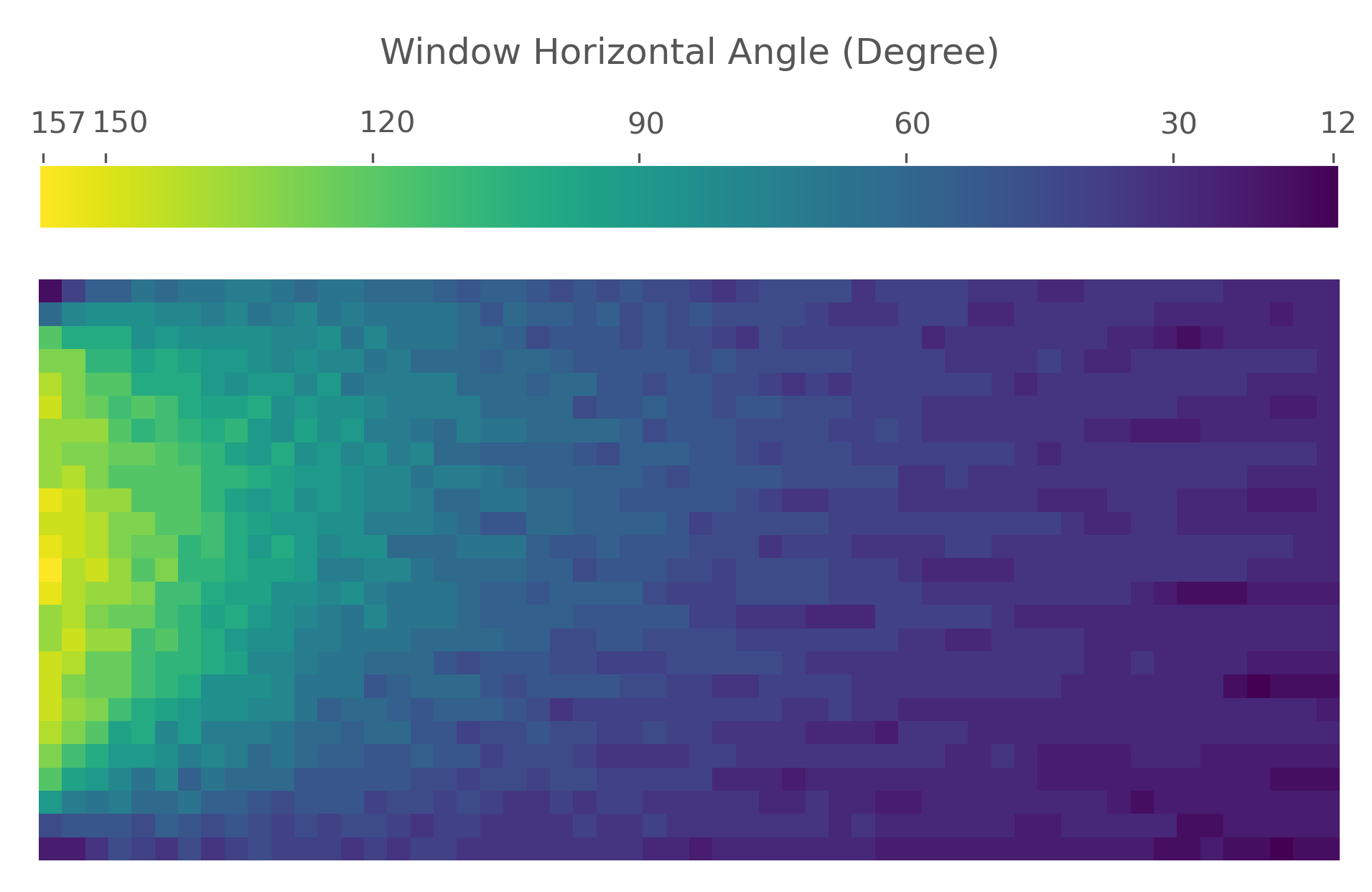}
	\caption{Values of horizontal sight angles across a grid of analysis points 1.7\,m high. The window is on the left side.}
	\label{fig:viewresult_angles}
	% \vspace{-4mm}
\end{figure}

Figure \ref{fig:viewresult_angles} shows the variation in horizontal sight angle values when calculated from different points in the room. As expected, the angle decreases as the distance from the window increases. The maximum angle found for the reference room is 158\textsuperscript{o} and the minimum is 11\textsuperscript{o}. If the point within the room with the minimum horizontal sight angle was to be considered for the evaluation, the room would not comply with any of the recommended performance levels. Given that this reference office room is characterized by a deep plan, this outcome was expected too.

\subsection{Exposure to sunlight}

The second set of results shows the ``Exposure to sunlight'' performance of the test scene. The direct view of the sun was evaluated for the whole period suggested by the standard, from the 1\textsuperscript{st} of February to the 21\textsuperscript{st} of March. The standard requires to choose only one day within this period for the evaluation of the cumulative number of hours of direct sunlight. In the analysed case, no major differences were found among different days, and the number of hours exceeded the threshold for the `High' performance level independently of the chosen day. 

\begin{figure}[!htbp]
	\centering
	\begin{subfigure}[a]{\linewidth}
		\centering
		\includegraphics[width=\linewidth]{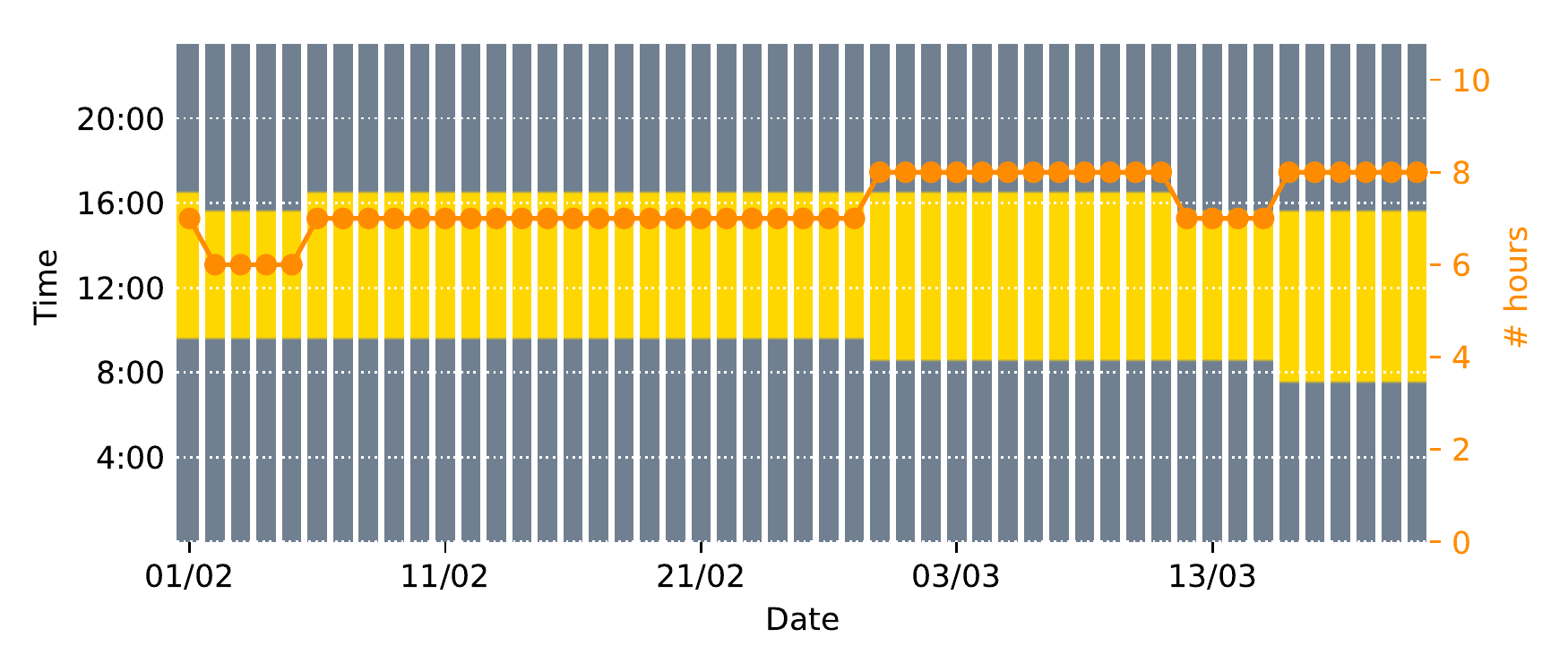}
		\caption{}
	\end{subfigure}
	\\
	\begin{subfigure}[b]{\linewidth}
		\centering
		\includegraphics[width=\linewidth]{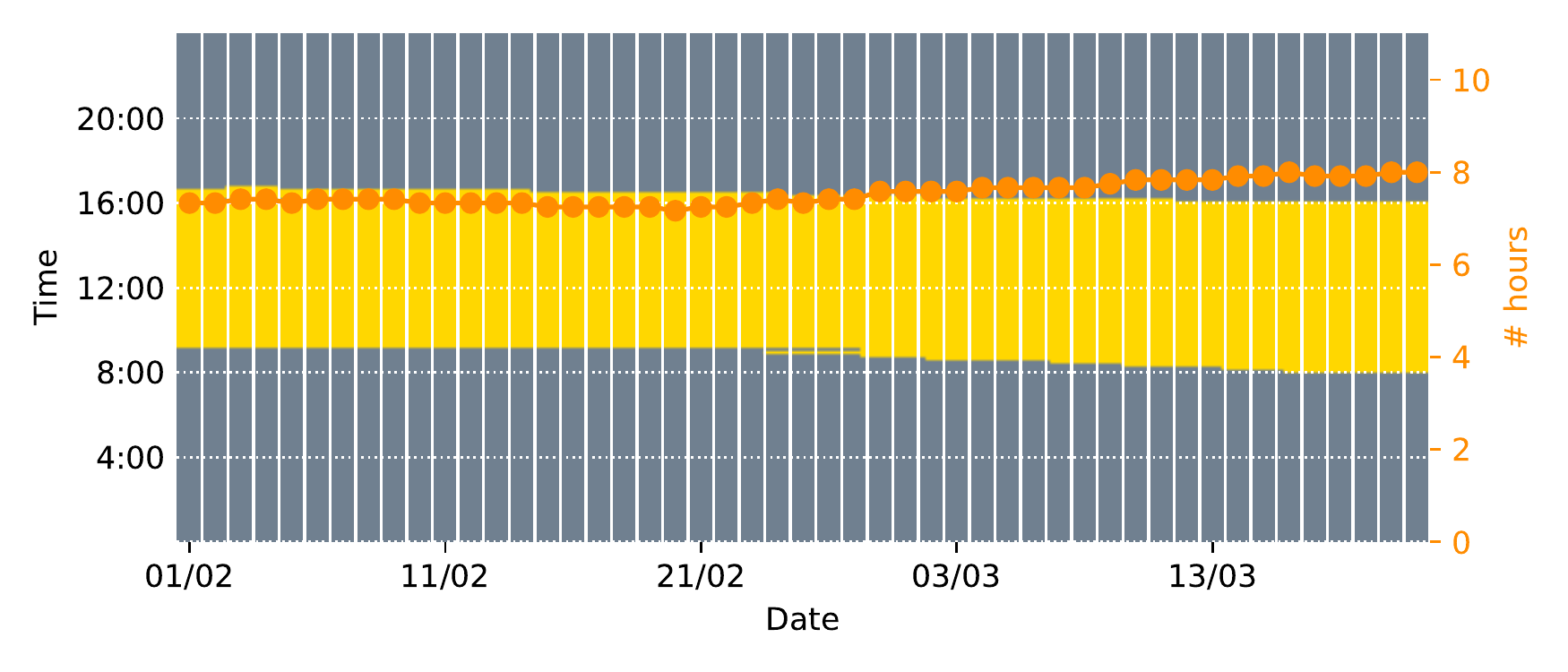}
		\caption{}
	\end{subfigure}
	\caption{Solar access at the evaluation point, within the period 01/02-21/03, for clear sky conditions. The plots show both the time at which direct sunlight access occurs (left \emph{y} axis) and the cumulative number of `direct sunlight hours' per day (right \emph{y} axis). Plot (a) shows results when using a 1 hour time step; plot (b) shows results for a 5 minute time step.}
	\label{fig:solarresult}
%	\vspace{-4mm}
\end{figure}

For this analysis, a decision on the timestep to adopt when defining sun positions had to be taken. In Figure \ref{fig:solarresult}, results obtained using a one-hour timestep are compared to those obtained using a five-minute timestep. In this case, a timestep of one hour led to a slight underestimation of exposure hours (-3\%). More analyses are required to assess the sensitivity of this performance criterion to the chosen timestep. In any case, specifying an exact value in the standard document would help reducing this uncertainty and making it more robust.

%\begin{figure}[!htbp]
%	\centering
%	\includegraphics[width=\linewidth]{img/EN_17037_sun_access_sunlight_hours_year}
%	\caption{Solar access at the evaluation point for the entire year, under clear sky conditions. The plot shows both the time at which direct sunlight access occurs (left \emph{y} axis) and the cumulative number of `direct sunlight hours' per day (right \emph{y} axis). The shaded portion on the left corresponds to the period indicated in the standard.}
%	\label{fig:solarresult_year}
%	%	\vspace{-4mm}
%\end{figure}

Besides considerations on detailed parameters within the ``Exposure to Sunlight'' criterion, considerations concerning the integration of this analysis with the other criteria in the standard document are important. Such a high incidence of direct sunlight should raise concerns about potential glare risk. To effectively inform designers of the need for shading devices and their optimal characteristics, the two criteria ``Exposure to sunlight'' and ``Protection from glare'' should be assessed in parallel. However, for such comparison to be possible, the two criteria should have more similar temporal and spatial requirements, as highlighted in the following section.

%\subsection{Variability}
%On a grid of points at the height of 1.2 meters above the floor the view out is computed and compared. We shall look at the range of values obtained from this analysis. The standard proposes to categorize the views towards the ground, landscape, the sky, emblematically labelled with Red, Green, and Blue colours. The hypothesis is that based on the given formulation of the standard a densely populated grid is likely to be biased towards giving a low score than a less densely populated grid, given the same location and the same view. The question is whether the test is robust enough to identify the differences as a human would care about. 

%\subsection{Comparison against standard examples}

\section*{Discussion}

The main scope of this work was to develop and present a tool that enables designers to check compliance with EN 17037 criteria. During its development, however, several inconsistent or missing variables were identified in the standard document. While it is true that the tool has been tested on a very limited number of geometries, the authors have sought to develop procedures that can be scaled and generalised to a wide range of scenarios. We discuss here the issues that we consider more relevant to ensure robust and consistent assessments of EN 17037 ``View out'' and ``Exposure to sunlight'' criteria, as well as general issues concerning all criteria in the standard. Leaving unclear and ambiguous definitions may lead to various interpretations and implementations by a single developer, hence to several different results and performance assessments. We argue that the implementation of additional parameters or thresholds should be the subject of discussion among the simulation and lighting community, ultimately resulting in modifications to the standard document itself. 

\subsection{Standardizing a standard} 
% The standard document addressed in this paper seems to require quite a few standardizations concerning definitions, computational methods, as well as required data inputs. In the following, we discuss each of such standardization issues in detail, but before going to the details, we shall recapitulate each issue with an example. 

% \begin{itemize}

%\subsection{Standard Definitions}

The suggested methods for ensuring the recommended levels of quality in the standard document are accompanied by a variety of 2D drawings and normative geometric methods that are supposed to help building designers attain the aforementioned qualities. While these drawings can be helpful in a didactic sense, they are not directly interpretable as unambiguous procedures for computing the values in question. The ambiguity arises specifically when translating the manual methods given in the standard into technical `recipes' that could be used for assessing 3D scenes. 

The absence of rigorous mathematical specification of the criteria increases the ambiguity of the computation procedures. As an example, when counting the visible layers in the view out criteria, a human's perception will discard miniscule visible patches of layers; but in the simulation environment, the rays will be likely to intersect with such objects. Then there has to be a defined threshold of solid angles in steradians, to identify the visibility of that layer unambiguously. As illustrated in Figure \ref{fig:viewresult_angles}, the problem of determining the number of visibility layers may lead to unexpected results if such computational details are not considered in the standard definitions.
	
	% an example of definitions to be standardized mathematically is the definition of the minimum solid view angle of objects for them to be counted in the view-out quality assessments. Suppose, for example, that in a high-resolution camera can capture a few pixels that pertain to the view of buildings some 12 kilometres away from a window; can a human perceive those objects? Most probably not. However, once this test performed computationally, the rays shot by the simulator will be likely to intersect with such objects. Then there has to be a defined threshold of solid angles in steradians, to be translated to the minimum number of hitting rays, corresponding to a standardized spherical tesselation determining the ray orientations. In the absence of such standard definitions, the problem of determining the number of 'building, landscape, and sky layers' will be simply ill-defined, as illustrated in  \ref{fig:viewresult_angles}.
	
%\subsection{Standard Data Models}

Lastly, it would be extremely valuable to consider the use of EN 17037 performance assessments within the wider field of urban and architectural design. This becomes particularly important for the ``View out'' assessment, which requires a fairly detailed description of the environment surrounding the building under analysis. The typical data models used by building simulation practitioners either lack semantic information (in the case of geometric models such as Wavefront OBJ or AutoCAD DXF made by Computer Graphics or Computer-Aided Design software) or their semantic layers do not correspond to the ones offered by the standard (in case of geometric-semantic data models such as gbXML and CityGML made by Building Information Modelling software or obtained from 3D City Information Models). In both cases, there is a need for a standardized semantic mapping between established data models and the semantic layers proposed by the standard (`ground', `landscapes', and `sky').

\subsection{Integration among EN 17037 criteria}

The two sections that were not investigated in this paper, concerned with ``Daylight sufficiency'' and ``Protection from glare'', proposed their own spatial and temporal aggregators. In the daylight sufficiency part, the annual evaluation option produces at least 8760 hourly values (of which about half are daylight hours) for each of the points in the analysis grid. Temporally, 50\% of the daylight hours should comply with the requirements. Spatially, two target performance values are defined: 50\% of the analysis grid should meet a target illuminance (e.g., 300 lx) and 95\% of the analysis grid should meet a minimum illuminance (e.g., 100 lx). In the protection from glare part, the annual evaluation option considers only occupied hours in a year (8:00 to 18:00, Monday to Friday), for a total of about 2600 values at each evaluated position. Of these hours, a maximum of 5\% should report a Daylight Glare Probability lower than a target value (e.g., 0.35). The evaluation position in the space should be decided by the designer and should correspond to the expected worse condition, i.e., close to the facade, with the sun shining in the occupant's field of view.

Similar aggregators could be easily devised also for the ``View out'' and ``Exposure to sunlight'' parts of the EN 17037 standard. The challenge lies with the selection of the most appropriate aggregator, as for it to be a meaningful representation of occupants' experience, more studies on human perception of view and sunlight would be required. Furthermore, more research is necessary for a more complete understanding of how the four criteria recommended by the standard interact with each other, and whether a design can achieve compliance with multiple criteria at once.

\section*{Conclusion}

This paper presented a computational toolset developed by the authors to assess the ``Exposure to sunlight'' and the ``View out'' criteria as defined by the EN 17037 standard ``Daylight in Buildings''. Besides illustrating the assumptions that had to be taken when interpreting the standard requirements, the present work aimed at suggesting more robust and unambiguous approaches for verifying compliance to the standard criteria. The analyses were carried out only for a single test scene, which is not necessarily representative of a typical office or an urban context. Yet, even the application of the standard requirements on a relatively simple room highlighted a number of issues that can affect the generalisation -- and hence the adoption -- of the standard. Among such issues, the major ones were: the absence of a defined point to assess the distance to obstructions; a definition of ``major obstructions''; a minimum solid angle threshold to account for view layers; and, a clearer procedure to choose the analysis day and timestep for the ``Exposure to sunlight'' assessment.

Furthermore, the paper discussed in more detail considerations about creating robust definitions, models and verification procedures for computational implementations; choosing result aggregators that more closely represent the human's perception of view and direct sunlight; and the need for integration among the four criteria covered by EN 17037. The latter aspect, i.e., the transition from simulation results to assessment labels invites further research in the direction of congruence with green building certifications such as BREEAM, LEED, and WELL. Furthermore, considering the demonstrated possibility of automated computational evaluation, a reconsideration of the evaluation paradigm from discrete labels to continuous scores would make the standard suitable for direct application in computational design workflows, especially for integration in objective functions for design optimization. 
\cite{CEN2018}

\section*{Acknowledgement}
The authors would like to acknowledge Mostapha Roudsari's help with understanding the Ladybug[$+$] and Honeybee[$+$] `recipes' structure. 

\section*{Supplementary Material}
The source code of the toolbox is available at the following repository: \href{https://github.com/shervinazadi/EN_17037_Compliance}{\url{https://github.com/shervinazadi/EN\_17037\_Compliance}}. Links to supplementary materials such as documentation, examples and online reproduction of the presented procedure are also available in the repository.

\AtNextBibliography{\small}
\printbibliography

\end{document}